\documentclass[%
 aip,
 amsmath,amssymb,
 reprint,%
]{revtex4-1}

\usepackage{graphicx}% Include figure files
\usepackage{dcolumn}% Align table columns on decimal point
\usepackage{bm}% bold math
\usepackage{color}
\usepackage[utf8]{inputenc}
\usepackage[T1]{fontenc}
\usepackage{mathptmx}
\usepackage{etoolbox}
\usepackage{physics}
\usepackage[separate-uncertainty=true]{siunitx}

\makeatletter
\def\@email#1#2{%
 \endgroup
 \patchcmd{\titleblock@produce}
  {\frontmatter@RRAPformat}
  {\frontmatter@RRAPformat{\produce@RRAP{*#1\href{mailto:#2}{#2}}}\frontmatter@RRAPformat}
  {}{}
}%
\makeatother
\begin{document}

\preprint{AIP/123-QED}

\title{Robust high-temperature atomic beam source with a microcapillary array}
% previous paper title
%Effusive Atomic Oven Nozzle Design Using an Aligned Microcapillary Array
% Force line breaks with \\
\author{Peter Dotti}\affiliation{ 
 $^{1)}$ Department of Physics, University of California Santa Barbara,\\ Santa Barbara, California 93106, USA%\\This line break forced with \textbackslash\textbackslash
}
\author{Xiao Chai}
\affiliation{ 
 $^{1)}$ Department of Physics, University of California Santa Barbara,\\ Santa Barbara, California 93106, USA%\\This line break forced with \textbackslash\textbackslash
}
\author{Jeremy L. Tanlimco}%
\affiliation{ 
 $^{1)}$ Department of Physics, University of California Santa Barbara,\\ Santa Barbara, California 93106, USA%\\This line break forced with \textbackslash\textbackslash
}
\author{Ethan Q. Simmons}%
\affiliation{ 
 $^{1)}$ Department of Physics, University of California Santa Barbara,\\ Santa Barbara, California 93106, USA%\\This line break forced with \textbackslash\textbackslash
}
\author{David M. Weld$^*$}%
\affiliation{ 
 $^{1)}$ Department of Physics, University of California Santa Barbara,\\ Santa Barbara, California 93106, USA%\\This line break forced with \textbackslash\textbackslash
}
 \email{weld@ucsb.edu.}

% \author{C. Author}
%  \homepage{http://www.Second.institution.edu/~Charlie.Author.}
% \affiliation{%
% Second institution and/or address%\\This line break forced% with \\
% }%

\date{\today}% It is always \today, today,
             %  but any date may be explicitly specified

\begin{abstract}
We present a new design for a directed high-flux high-temperature atomic vapor source for use in atomic physics experiments conducted under vacuum.  An externally heated nozzle made of an array of stainless steel microcapillaries produces a collimated atomic beam.  Welded stainless steel construction allows for operation at high source temperatures without exposing delicate conflat vacuum flanges to thermal stress, greatly enhancing robustness compared to previously published designs.  We report in operando performance measurements of an atomic beam of lithium at various operating temperatures.
\end{abstract}

\maketitle

% \begin{quotation}
% The ``lead paragraph'' is encapsulated with the \LaTeX\ 
% \verb+quotation+ environment and is formatted as a single paragraph before the first section heading. 
% (The \verb+quotation+ environment reverts to its usual meaning after the first sectioning command.) 
% Note that numbered references are allowed in the lead paragraph.
% %
% The lead paragraph will only be found in an article being prepared for the journal \textit{Chaos}.
% \end{quotation}

\section{\label{sec:introduction}Introduction}

Atomic and molecular beams have been a critical component of physics experiments for more than a century\cite{SternGerlach_1922}, and several reviews have detailed the tradeoffs and considerations involved in source design.\cite{Lucas_text,Ross.Sonntag.1995} Typically, sources comprise a heated reservoir and a nozzle that shapes and expels a directed beam of atomic vapor. We present a new source design that produces a high-flux and well-collimated atomic beam, is straightforward to assemble, and is optimized for operation at high temperatures without risk of vacuum leaks. 

\begin{figure}[b!]
\includegraphics[width=3.37in]{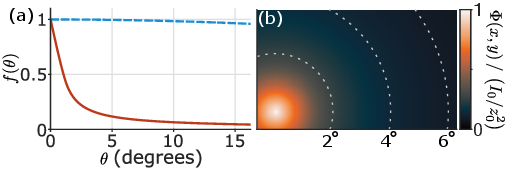}% Here is how to import EPS art
\caption{\label{fig:single_capillary_flux} (a) Theoretically predicted normalized angular distribution of the flux from a nozzle with a circular aperture in a thin plate (blue dashed) compared to that of a single microchannel with the same radius $a=0.0546$ mm and length $L=5$~mm (red solid). (b) Predicted flux through a plane at $z_0$ downstream of the single microchannel nozzle.  Dotted white circles correspond to distances of $z_0\sin(\theta)$ from the beam centerline where $\theta$ is indicated along the x axis. }
\end{figure}

The design we present utilizes a microchannel array.\cite{Gordon.Townes.1955,King.Zacharias.1956,Helmer.Sturrock.1960,Giordmaine.Wang.1960,Johnson.Pritchard.1966,Ross.Sonntag.1995,Schioppo.Tino.2012,lucas2013atomic,Bowden.Madison.2016,RuwanRSI,Li.Raman.2019,Li.Raman.2020,Wodey.Schlippert.2021}  
A key figure of merit is the radiant flux $\dv*{\dot{N}}{\Omega}$, which describes the number of particles per second per steradian as a function of direction.  An azimuthally symmetric  flux can be described by
$
\dv*{\dot{N}}{\Omega} = I_0 f(\theta), 
$
using spherical coordinates $(\theta,\phi)$ and taking $\theta=0$ to correspond to the beam axis and $f(0) = 1$.  The simplest  nozzle consists of a small aperture in a thin plate,\cite{Ramsey_text} with $f(\theta)=\cos(\theta)$. Improved collimation can be achieved by using an array of long narrow channels (hexagonally packed segments of stainless steel hypodermic tubing in our design).
Fig.~\ref{fig:single_capillary_flux}(a) compares the predicted normalized angular distribution of the flux $f(\theta)$ for a nozzle made from a circular hole in a thin plate to that from a single microchannel\cite{Beijerinck} of the same radius.  The theoretical model assumes operation in the transparent regime (justified in the Supplementary Material) so that $I_0$ is the same in both cases.  The angular distribution of the flux is much more strongly peaked for the channel, with 3\% of the total flux directed within $\ang{3}$ of the centerline.  This is a factor of 22 increase in fractional flux through the $\theta<\ang{3}$ region compared to the thin plate case.   Fig.~\ref{fig:single_capillary_flux}(b)  shows the predicted flux $\Phi$ from the channel through a planar target placed perpendicular to the $\theta=0$ axis at a distance $z_0$ from the source.  

While microchannel nozzles are thus more efficient, to attain sufficient total flux while maintaining this efficiency it is necessary to assemble a large, defect-free, and uniformly-directed array of many such channels. Furthermore, to avoid clogging, the array should be heated to a temperature well above that of the oven itself (which for a low vapor pressure species like Li or Sr can be $\SI{500}{\degreeCelsius}$ or more, exceeding the temperature rating of common vacuum components like CF flanges). The design we present here addresses both of these challenges, greatly enhances robustness against vacuum leaks, and offers improved ease of assembly and manufacturability over previous designs.\cite{RuwanRSI}  The stainless steel construction allows for operation with a variety of sources\cite{Ross.Sonntag.1995} and the unheated and passively air-cooled CF flange allows for nozzle temperatures in excess of $\SI{550}{\degreeCelsius}$ with greatly reduced risk of vacuum leaks.

\begin{figure*}
\includegraphics[width=7in]{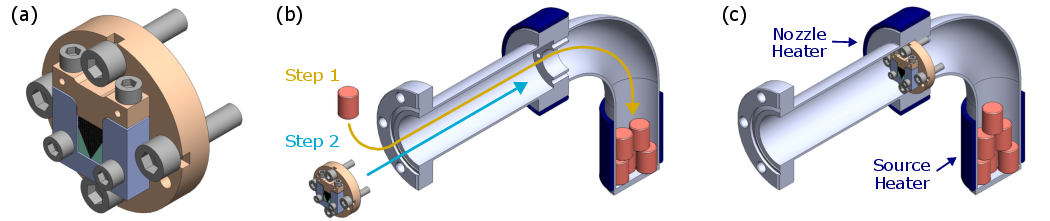}%
%{2022OvenSection.PDF}%Here is how to import EPS art
\caption{\label{fig:vacuumChamberSegment} (a) Schematic of the assembled nozzle. False coloring indicates different components.  (b) A cross sectional view of the vacuum chamber segment, showing partially loaded source material as red cylinders.  Once loaded (Step 1), the assembled nozzle is fixed at the internal mounting point in the larger-outer-diameter segment (Step 2). Note that no vacuum flange is present in this welded mounting segment, enabling safe heating to very high temperatures.  (c) Oven assembly after installation of the nozzle.  The locations of nozzle and source heaters are indicated in dark blue. Threads on bolts and in tapped holes are not shown.}
\end{figure*}

\section{\label{sec:designs}Description of design and assembly}

The atomic source has three main components: the nozzle, shown in Fig.~\ref{fig:vacuumChamberSegment}(a), the oven chamber, and the external heating elements.  The three components are constructed separately.  During assembly, the source material is loaded into the cup-like reservoir region through a 0.77 inch diameter hole in the nozzle mounting segment, after which the nozzle is attached to the internal mounting segment as shown schematically in Fig.~\ref{fig:vacuumChamberSegment}(b). Threaded blind holes on the cover plate of the nozzle are used as an aid in the installation process: a long threaded rod can be screwed into them to act as a handle during installation. These assembly steps must be carried out in a suitable environment, such as an argon-filled glovebag if the source is reactive in air. After nozzle installation, the vacuum chamber segment is attached to the larger experimental vacuum chamber and evacuated.  The heating elements, shown as dark blue bands in Fig.~\ref{fig:vacuumChamberSegment}(b) and Fig.~\ref{fig:vacuumChamberSegment}(c), are then attached to the exterior of the vacuum chamber segment, insulation is wrapped around the oven, and the vacuum chamber is baked if necessary. The fully assembled atomic source is shown without insulation in Fig.~\ref{fig:vacuumChamberSegment}(c).

%\subsection{\label{sec:vacChamberSegment}Vacuum chamber segment}

The vacuum chamber segment shown in Fig.~\ref{fig:vacuumChamberSegment}(b) is a welded assembly of standard vacuum chamber elements and one custom piece that serves as an internal mounting point for the nozzle.  The custom part is machined from stainless steel using standard techniques.  These components are joined with vacuum-tight gas tungsten arc welding.  A fully assembled vacuum chamber segment can be ordered from a vacuum chamber manufacturer.\cite{ANCORPNote} In this design the cup region and the nozzle are directly heated and insulated, but the CF flange is not.  Thus the source temperature is limited only by what the steel chamber and welds are rated to allow and by the temperature that the passively air-cooled flange reaches due to conductive heat transfer from the oven. For a nozzle temperature of $\SI{550}{\degreeCelsius}$ the CF flange on our experiment reaches only $\SI{75}{\degreeCelsius}$ measured externally, well within the safe operating range.

\begin{figure}[b]
\includegraphics[width=3.37in]{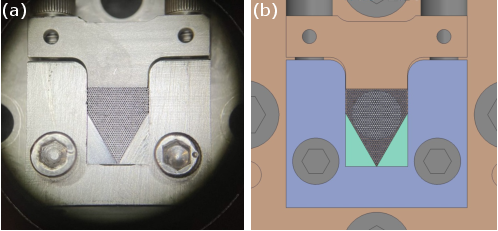}% Here is how to import EPS art
\caption{\label{fig:nozzleImage} (a) Photograph of assembled nozzle.  (b) CAD rendering of the nozzle from a similar perspective as (a). The cover plate is shown in bronze, the U-shaped cradle is shown in blue, and the two wedge pieces are shown in green.  Standard 4-40 and 8-32 bolts and the microcapillaries are shown in gray.}
\end{figure}

%\subsection{\label{sec:nozzle}Nozzle}
The central component of the nozzle is a hexagonal array of approximately 800 parallel microcapillaries clamped together by several custom machined parts.  The microcapillaries were ordered\cite{microgroupNote} to have lengths of $\SI{5\pm0.025}{\milli\meter}$.  The nominal outer diameter was $\SI{0.21}{\milli\meter}$ (8.3~thou) with a permitted range of $\SI{0.203}{\milli\meter}$ (8.0~thou) to $\SI{0.215}{\milli\meter}$ (8.5~thou).  The nominal inner diameter was $\SI{0.11}{\milli\meter}$ (4.3~thou) with a permitted range of $\SI{0.090}{\milli\meter}$ (3.5~thou) to $\SI{0.127}{\milli\meter}$ (5.0~thou).  The microcapillaries were made of 304 stainless steel because 316 stainless was not available.   An image of an assembled nozzle is shown in Fig. \ref{fig:nozzleImage}(a) and a corresponding CAD rendering with false color components is shown in Fig. \ref{fig:nozzleImage}(b). Dimensioned drawings of all custom machined parts of the designs are presented in the supplementary material along with an assembly fixture to aid in the stacking of the microcapillaries into the channel of the the nozzle. 

\begin{figure*}[t!]
\includegraphics[width=7in]{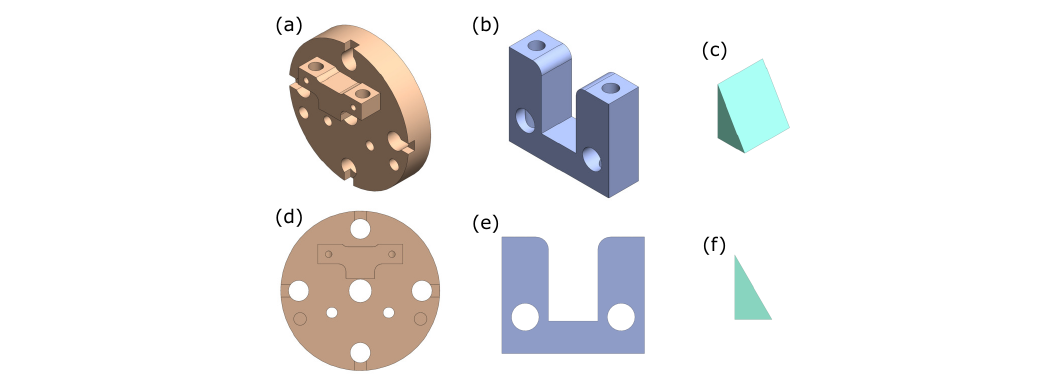}% I needed to make this a pdf so that arXiv handled the upload properly (it messed with the supplemental material strangely.)  It can be eps otherwise
\caption{\label{fig:nozzleComponents} Schematic illustrations of the the individual nozzle components.  An isometric view of the cover plate (a), the U-shaped cradle (b), and one of the two wedges (c) are shown.  The components in the same order are shown as viewed along the beam axis in (d), (e), and (f).  Threads are suppressed here; threading is indicated in the dimensioned drawings in the Supplementary Material.}
\end{figure*}

The hexagonal lattice structure of the microcapillary, which is necessary to align all the capillary axes and to avoid large gaps due to packing defects which can dominate the nozzle conductivity, is ensured by a channel with sides sloped to create a $\ang{60}$ opening.  The $\ang{60}$ angled sides, along with the sharp corner at the bottom of the triangular part of the array, together create a boundary condition which enforces hexagonal packing of the microcapillaries and preserves it once they are clamped into place.  This design differs from previous designs\cite{RuwanRSI} in the use of wedge pieces to create the $\ang{60}$ channel.  These wedge pieces and the microcapillaries are held in place by  compression between the T-shaped extrusion on the cover plate component (Fig.~\ref{fig:nozzleComponents}(a) and (d)) and the square bottom U-shaped cradle  (Fig.~\ref{fig:nozzleComponents}(b) and (e)).  Using wedges circumvents the practical difficulty of machining an interior corner with a radius comparable to the outer radius of the microcapillaries.  In the previous design such a corner was made using wire electrical discharge machining (EDM).  Wire diameters of $\SI{0.25}{\milli\meter}$ were used in previous designs for microcapillaries with nominal diameters of $\SI{0.21}{\milli\meter}$; this resulted in a slight defect in packing near the corner which sometimes complicated assembly.  In the designs presented here, the requisite sharp corner is created from sharp exterior corners on the wedges, that are easily created on an ordinary milling machine.  However, this design does require that a few critical features of the nozzle be machined to a precision that is significantly less than the $\SI{0.21}{\milli\meter}$ microcapillary diameter.  These features are the width of the wedges, the interior width of the U-shaped cradle, and the width of the T-shaped extrusion on the cover plate. For these lengths a tolerance of $\SI{\pm 0.025}{\milli\meter}$ (1~thou) is sufficient.  This is a demanding tolerance, but one that is commonly achievable by professional machinists on parts of this size.

The U-shaped cradle in Fig.~\ref{fig:nozzleComponents}(b) and (e) holds the two identical wedges shown in Fig.~\ref{fig:nozzleComponents}(c) and (f) to set the boundaries for the array of microcapillaries. The U-shaped cradle has sharp interior corners that can be easily machined using a square end mill from above. Another departure from previous designs is that the sloped sides transition to vertical once the microcapillaries are stacked above the wedges.  This allows for the number of microcapillary rows to be variable and enables strong compression to be applied even if the microcapillary array is compliant under the applied forces.  This choice gives rise to a ragged ``armchair'' boundary of the microcapillary array along these vertical edges and minor defects at the corners, but does not significantly impair proper packing of the microcapillaries in the important central region.

The cover plate shown in Fig.~\ref{fig:nozzleComponents}(a) and (d) is the largest component of the nozzle.  Once the nozzle is assembled, the four vented 8-32 through holes near the perimeter are used to mount the nozzle inside the vacuum chamber segment.  The central through hole in the cover plate determines which capillaries the atoms can pass through.

%\subsection{\label{sec:heating}Heating elements and insulation}

To heat the nozzle and source, we attach two band heaters\cite{TEMPCONote} to the exterior of the oven.  One of the heaters, with $\SI{57}{\milli\meter}$ (2.25~inch) internal diameter and $\SI{25}{\milli\meter}$ (1 inch) length, is clamped around the welded nozzle section.  The other, with a $\SI{38}{\milli\meter}$ (1.5 inch) internal diameter and a length of $\SI{51}{\milli\meter}$ (2 inches) is clamped around the end of the source cup.  As in previous designs, we operate with the nozzle $\SI{50}{\degreeCelsius}$ hotter than the reservoir to prevent clogging of the nozzle caused by migration of the atomic sample into the microcapillaries.

We insulate the source with a several-inch-thick layer of superwool sheets each enveloped in aluminum foil.  The majority of the vacuum chamber segment is wrapped in insulation with the exception of the CF flange and approximately 1 inch adjacent to the flange. The lack of insulation near the flange enables passive cooling, keeping the flange well below its maximum operating temperature.  At the highest tested nozzle temperature of $\SI{550}{\degreeCelsius}$ (cup temperature $\SI{500}{\degreeCelsius}$) we measure that the external flange temperature  reaches only $\SI{75}{\degreeCelsius}$ as measured with a J-type thermocouple attached to the exterior.  In a decade of experience with the previous design, in which the nozzle is embedded within a CF flange, we found that the heated flanges could eventually leak if the flange was repeatedly temperature cycled, and that the CF bolts could degrade due to extended exposure to elevated temperatures.  This vulnerability is essentially eliminated in this design.

We note that assembly of the nozzle can be challenging.  In particular, manipulation of the tiny microcapillaries to arrange them into an array takes some practice, even with the aid of the construction scaffold described in the supplementary material.  We recommend that the nozzle be well illuminated and viewed under a stereo microscope during assembly, with the nozzle propped up at an angle.  Tweezers can be used to set the microcapillaries into place.  Once several rows of capillaries have been set, assembly can be made more efficient by dropping a few dozen microcapillaries onto the top of the partial array and then carefully agitating the top few layers with the tweezers until they ``anneal'' into place.  One must take care not to shift the wedges in the process or microcapillaries can become lodged beneath or behind them.  During assembly, the microcapillaries should be placed on a conductive grounded plate so that they do not accumulate static charge and repel each other.  The microcapillaries may nonetheless stick and repel slightly from magnetization.  Once enough microcapillaries have been stacked, the base plate should be carefully and securely clamped from above before the construction scaffold is removed.  %One can gently push on the end of the microcapillaries to make sure the clamping force is sufficient.

\section{\label{sec:experiments}Measured performance}

\begin{figure}[th!]
\includegraphics[width=3.375in]{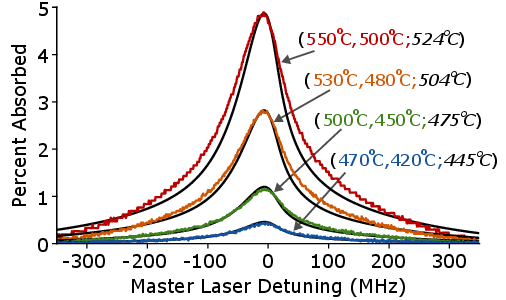}% Here is how to import EPS art
\caption{\label{fig:flux_absorption} Experimental measurement of the atomic beam.  Colored lines show absorption of a probe laser as a function of detuning at different nozzle temperatures.  Black lines indicate  theoretical predictions (see text for details.)  The temperatures are indicated in the form ($T_{\rm n}$,$T_{\rm s}$;$T_{\rm m}$), where $T_{\rm n}$ and $T_{\rm s}$ are the nozzle and source cup temperature, respectively, and $T_{\rm m}$ is the temperature used to generate the theory curve.  The measured conflat flange temperatures were $\SI{74.8}{\degreeCelsius}$, $\SI{72.5}{\degreeCelsius}$, $\SI{63.2}{\degreeCelsius}$, and $\SI{54.6}{\degreeCelsius}$.}
\end{figure}

An atomic beam source of the design we describe was installed on our ultracold lithium experiment, and the resulting flux of $^7$Li was measured at various temperatures.  A weak probe laser beam of $\SI{3.7}{\micro\watt}$ total power with approximately a $\SI{1}{\milli\meter}$ diameter was used to detect the atomic flux.  This linearly polarized probe beam reached the atoms after exiting a single mode fiber whose output was mounted on a translation stage so that the probe beam could be centered on the atomic beam.  The probe laser beam was generated by superimposing a ``cycler'' laser beam and a ``repump''  laser beam generated from the same master laser.  The resulting probe beam had two sharp frequency components of equal power,  nearly resonant with known detunings from the $\ket{2\,^2\mathrm{S}_{1/2};F=2} \rightarrow \ket{2\,^2\mathrm{P}_{3/2};F=3}$ and the $\ket{2\,^2\mathrm{S}_{1/2};F=1} \rightarrow \ket{2\,^2\mathrm{P}_{3/2};F=2}$ hyperfine transitions of $^7 \mathrm{Li}$. We measured the amount of probe beam light absorbed as it passed through the atomic beam as a function of the detuning of the master laser from the $\mathrm{D}_2$ saturated absorption peak. The detuning of the master laser was determined from a simultaneously measured spectroscopy signal.  This measurement was repeated for several atomic beam source temperatures. A selection of the results are shown in Fig.~\ref{fig:flux_absorption}.

%When the master laser was set to its lock point, these two components were near resonant with two 2$\,^2$S$_{1/2}\ \rightarrow 2\,^2$P$_{3/2}$ (D$_2$) hyperfine transitions of the $^7 \mathrm{Li}$.  Using $F$ and $F'$ to denote the S state and the P state hyperfine quantum numbers, respectively, one of the components of the probe beam excites atoms from the $F=2$ to $F'=3$ states, and the other from the $F=1$ to $F'=2$ state when the master laser is locked, with known detunings from exact resonance.

To more precisely interpret the absorption data, we ignore interatomic interactions and model the distribution of atoms emitted from each capillary in the nozzle using the angle-dependent flux  shown in Fig.~\ref{fig:single_capillary_flux}, using the Maxwell-Boltzmann distribution of particle flux for the velocity distribution.\cite{Beijerinck}  We can then calculate how much the probe beam would be attenuated as it passes through the modeled atomic beam.  The experiment is very sensitive to small deviations of the probe beam angle from perpendicular to the atomic beam axis.  Such tilts must be considered in our model as they distort the ideal signal and change the laser detuning at which maximum absorption occurs.  However, small translations of the probe beam have negligible effect.   We compared various model parameters to determine which beam tilts and source temperatures best fit the data, prioritizing matching the absorption peak of the model to the data.  We deduced that the probe beam had a tilt of $\ang{0.75}\pm\ang{0.10}$ relative to normal incidence for all data shown, and we fitted source temperatures to within $\SI{\pm 1}{\degreeCelsius}$.  The best matching source temperature is determined in this way because the nozzle heater and cup heater operate at unequal temperatures. The theoretical modeling procedure is described in greater detail in the supplementary material.

In Fig.~\ref{fig:flux_absorption}, each model is shown as a black curve and can be used to deduce the radiant flux of $^7$Li from a single capillary.  According to the models with the determined temperatures the single capillary peak radiant flux $I_0$ in order of decreasing temperature for each curve in Fig.~\ref{fig:flux_absorption} is 89.5, 50.0, 20.4, and 7.5 in units of $10^{12}$ $^7$Li atoms per second per steradian, with an uncertainty of about 5\%.  If one is far enough from the nozzle to ignore the $\SI{5}{\milli\meter}$ diameter of the nozzle,  the radiant flux of $^7$Li from the entire nozzle is about (475)$I_0^\text{Li7}$, since the number of capillaries that are less than half obstructed by the cover plate is 475.  The source has been in continuous use for 15 months without any clogging or vacuum issues. We operate our lithium BEC machine at a source temperature of $\SI{475}{\degreeCelsius}$ and find the resulting fluxes to be sufficient to create Bose-Einstein condensates of more than a million atoms. 

\section{CONCLUSION}
We have presented an updated design of a cost-effective atomic beam source, suitable for cold-atom experiments under ultra-high vacuum. Constructed from welded stainless steel components, the oven operates at high source temperatures while maintaining the CF flange at much lower temperature, which significantly enhances the long-term operational lifetime. We calibrate the performance of the source in our $^7$Li setup using absorption spectroscopy. The resulting atomic beam is well-collimated with a total flux up to $3.81 \times 10^{15}$ atoms per second and a radiant flux up to $4.25\times 10^{16}$ atoms per second per steradian along the beam axis.  %The details of the design and assembly are given for the convenient reproduction of the source. 

\begin{acknowledgments}
The authors thank Eber Nolasco-Martinez and Xuanwei Liang for experiment assistance and acknowledge support from the the National Science Foundation (OMA-2016245 and 2110584), the Army Research Office (W911NF2210098), the UC Noyce Initiative, and the Eddleman Center for Quantum Innovation.
\end{acknowledgments}

\section*{Data Availability Statement}
The data that support the findings of this study are available from the corresponding author upon reasonable request.

\section*{Supplementary Material}

In the supplementary material of this manuscript, we justify the use of the transparent regime analysis, describe the modeling of the flux measurement in detail, and provide dimensioned technical drawings of the vacuum chamber segment and nozzle components.

\bibliography{aipsamp}% Produces the bibliography via BibTeX.

\end{document}

\maketitle
%\begin{center}
%{\Large Supplementary Material to \\
%Robust high-temperature atomic beam source with a microcapillary array}
%Peter Dotti, Xiao Chai, Ethan Q. Simmons, Jeremy L. Tanlimco, and David M. Weld
%\end{center}

\section{Transparent Regime Justification}

The output of a long narrow nozzle is often modeled using an analytical expression which neglects interatomic collisions \cite{Beijerinck}.   In reality, many experiments including ours operate the nozzle close to the boundary between this transparent regime and the opaque regime where collisions are important. In our case this leads to at most a few percent error, an assertion we justify here.

The transparent regime occurs when it is negligibly probable that an atom entering a microchannel will collide with another atom before it either (A) exits the nozzle or (B) collides with a wall of the microchannel.  In this regime, one can ignore any reduction in the peak radiant flux $I_0$ at $\theta=0$ due to interatomic collisions.  Such collisions otherwise increase the radiant flux at large values of $\theta$; intuitively this can be thought of as resulting from a decreased effective length of the capillary since atoms appear to originate from the point in the microchannel where collisions create an approximate Boltzmann distribution of velocities \cite{Beijerinck}.  To show that our nozzle was operated in the transparent regime, we will compare the mean free path $l_{\text{mf}}$ of an atom in the source oven to the length $L = \SI{0.5}{\centi\meter}$ of a cylindrical microchannel of the nozzle.

To determine a lower bound on $l_{\text{mf}}$, we first calculate the vapor pressure of the lithium atoms in the source assuming the source to be at a uniform temperature of $\SI{550}{\degreeCelsius}$, the hottest tested temperature of the nozzle.  This pressure is $P=\SI{1.91}{\pascal}$ \cite{Mondal_Antoine_Params}.  The ideal gas law implies an atom density $n_{\mathrm{s}}=\SI{1.68e20}{\meter^{-3}}$ of all isotopes of lithium in the source.  Finally, using the van der Waals diameter of a lithium atom $d_{\mathrm{vdW}}=\SI{3.62e-10}{\meter}$ \cite{doi:10.1021/jp8111556}, we get
\[
l_{\text{mf}} = \frac{1}{\sqrt{2}\,\pi n_\text{s} d_\text{vdW}^2} = \SI{1.02}{\centi\meter}
\]

Thus, we can confidently say that $l_{\text{mf}}/L \ge 2$ throughout this work.  For the rest of this section, we will use the more realistic source temperature of $\SI{525}{\degreeCelsius}$ as determined from the analysis of flux measurements described in the main text.  For this source temperature $l_{\text{mf}}/L \approx 4$.

As the particle density increases so that one begins to leave the transparent regime and enter the opaque regime, the reduced peak radiant flux $I_0^*$ can be approximated as \cite{Beijerinck,Giordmaine_and_Wang}
\[
I_0^* = I_0 \sqrt{\frac{\pi}{2 n^*}\,}\text{erf}\left( \sqrt{n^*/2} \right)
\]
where $n^* = L/l_\text{mf}$ and $I_0$ is the calculated peak radiant flux assuming no atom collisions.  Plugging in $n^*=L/l_{\text{mf}}=1/4$, one finds 
\[
I_0^* = (0.96)I_0
\]
This 4\% error is not very important for the purposes of assessing the quality of our nozzle.  The approximations made in using an analytical form for the radiant intensity rather than performing a Monte Carlo simulation of atom trajectories already lead to a few percent error underestimate of $I_0$ \cite{Beijerinck}.  Given the exponential growth in source vapor density with source temperature, such errors are of little practical relevance for the applications considered in this paper.  

\section{Flux Measurement and Model Details}

To theoretically model the measured fluxes shown in Fig.~5 of the main text, we must first determine the absorption cross section of $^7$Li atoms leaving the nozzle.  Let us denote states of $^7$Li in the 2\,$^2$S$_{1/2}$ ground state manifold as $\ket{g,f,m_f}$ and states in 2\,$^2$P$_{3/2}$ excited state manifold as $\ket{e,f,m_f}$, where $f$ and $m_f$ denote the hyperfine quantum number and corresponding total spin quantum number, respectively. All other states of lithium can be safely neglected.  To describe atomic states, we denote the density matrix
\[
\boldsymbol{\rho} = \sum\rho_{\alpha,f,m_f;\alpha',f',m_f'} \ket{\alpha,f,m_f}\!\bra{\alpha',f',m_f'}
\]
with the sum over $\alpha,\alpha'\in\{g,e\}$ and over all $f$, $f'$, $m_f$, and $m_f'$ values. The temperature of the source oven lets us safely approximate the initial $^7$Li atomic states by $\rho_{g,f,m_f;g,f,m_f} = 1/8$ for all values of $f,m_f$ and all other matrix elements zero, i.e.\! equal incoherent population of all 2\,$^2$S$_{1/2}$ states. 

We approximate the absorption cross section for a laser  that is near resonant with the 2\,$^2$S$_{1/2}$ to 2\,$^2$P$_{3/2}$ transition by
\[
\sigma(\bar{\omega}) = \frac{\hbar \omega}{I} \sum \frac{|\Omega(g,f,m_f;e,f',m_f')|^2}{\Gamma\left[\!1 +  [2\Delta(\bar{\omega},f,m_f,f',m_f')/\Gamma]^2\!\right]}\rho_{g,f,m_f;g,f,m_f}
\] 
where $\omega$ ($\bar{\omega}$) denotes the angular frequency of the laser beam in the lab (atom) frame, $I$ denotes the laser beam intensity, $\Gamma$ denotes the linewidth of the transition, $\Delta(\bar{\omega},f,m_f,f',m_f')$ denotes the detuning of the laser beam from resonance with the $g,f,m_f$ to $e,f',m_f'$ transition in the atom's rest frame, and $\Omega(g,f,m_f;e,f',m_f')$ is a hyperfine Rabi frequency \cite{Steck_notes}.  This assumes a weak probe beam, so that the saturation parameter is small. In our experiment, the saturation parameter is about $0.038 \ll 1$. This allows us to ignore optical pumping for the duration that the atoms interact with the probe beam and to assume that the matrix elements $\rho_{e,f,m_f}$ that determine the spontaneous decay rate reach small equilibrium values proportional to $I$ with negligible-amplitude Rabi oscillations.

Recall that our probe beam is composed of two different sharply peaked frequency components.  To treat this, we calculate the absorption cross section for each of the two frequencies and then we average the cross sections weighting by the intensity of each frequency component.  This yields the full effective absorption cross section $\tilde{\sigma}(\tilde{\omega})$, where $\tilde{\omega}$ is a reference frequency for the two frequency component probe beam.

Let $\mathbf{r}_c$ denote the position vector for the exit of a single capillary. We can describe the density and velocity distribution of atoms from this capillary at position $\mathbf{r}$ by the function $n(\mathbf{r}_\delta,v)$, where we let $\mathbf{r}_\delta \equiv \mathbf{r}-\mathbf{r}_c$. The velocity vector of atoms at position $\mathbf{r}$ from the capillary will be $\mathbf{v}=v\hat{\mathbf{r}}_\delta$, where $\hat{\mathbf{r}}_\delta =\mathbf{r}_\delta/|\mathbf{r}_\delta| $.  
Next, let us consider the path of a narrow probe laser beam described by the line $s\hat{\mathbf{k}} + \mathbf{r}_0$ where $s$ parameterizes the distance from the laser beam source $\mathbf{r}_0$. The optical density OD of the entire atomic beam with a probe beam reference frequency $\tilde{\omega}$ is
\[
\text{OD} = \sum_{\mathbf{r}_c} \int\! \mathrm{d}{s} \int\! \mathrm{d}{v}\ n(\mathbf{r}_\delta,v)\ \tilde{\sigma}(\tilde{\omega}-[\,\hat{\mathbf{k}}\cdot\hat{\mathbf{r}}_\delta]v )
\]
where the sum is over all capillary positions.  The percent transmitted is simply $100\exp(-\text{OD})$.  In Fig.~5 of the main text, the theoretical curves were created by computing the percent absorbed for every laser frequency value recorded in the data.

In our experiments, atoms leaving the capillaries at wide angles will hit the relatively cool vacuum chamber walls, where we assume they stick and are lost.  We use the maximum angle for which no collision occurs to set the limits of the $s$ integral.  As a simplification, we calculate the limits assuming all atoms originate at the center of the nozzle, which we confirmed negligibly affects the results.

A few other considerations were made in matching the model to the data.  The tilt of the probe beam is accounted for by changing $\hat{\mathbf{k}}$.  For computational speed, the velocity integration was tabulated beforehand, and an interpolation of this tabulation was used in evaluations of the $s$ integrals in the sum.  $\mathbf{r}_0$ for the probe beam was varied within the error of the beam position, but found to be unimportant. There remained some discrepancy between the theory and experiment in Fig.~5 of the main text, which we speculate is primarily caused by the theoretical model not fully capturing the details of the atomic trajectories away from the nozzle.  A more accurate model might be achieved with Monte Carlo simulations and a fuller accounting of interatomic collisions.  We note that we neglected the finite width of the probe laser beam in our analysis, which was found to be important in related work \cite{Li.Raman.2020}, but numerical tests justified our neglecting the finite width of the beam in our case which can be explained by the circular aperture of our nozzle being large compared to the probe beam.  Also, the probe beam intensity was not stabilized during the frequency scan, so its intensity varied a small amount that was important compared to the percent scale absorption signal.  To correct for this, we used regions where nearly all of the laser light was transmitted (not shown in Fig.~5) to make a quadratic fit that was subtracted from the data.  This probably introduces a small signal distortion at nozzle temperatures above $\SI{500}{\degreeCelsius}$, likely causing a slightly lower observed absorption in the data at the higher detuning shown in Fig.~5 of the main text.

\bibliographystyle{unsrt}
\bibliography{aipsamp.bib}

\section{Dimensioned Drawings of Components}

\begin{figure}[th!]
\begin{center}
\includegraphics[width=2.75in]{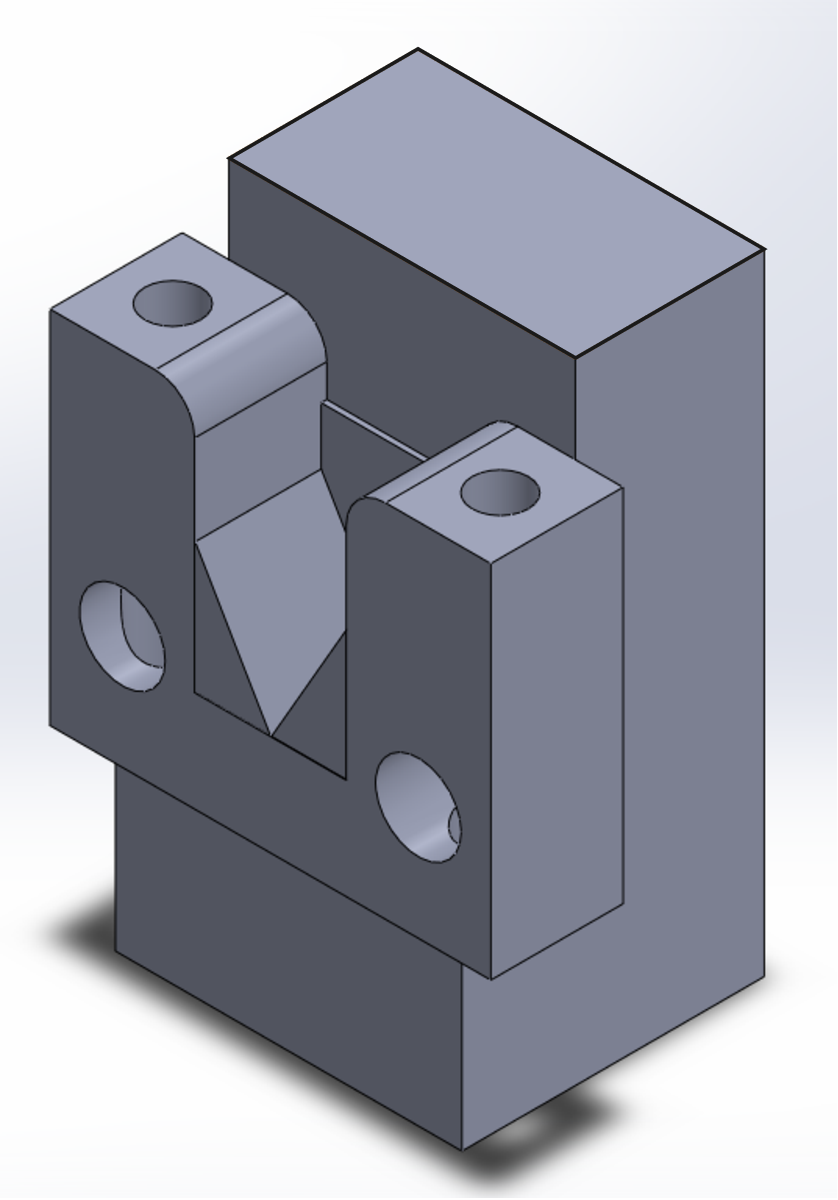}% Here is how to import EPS art
\caption{\label{fig:construction_scaffold_assembled} The construction scaffold shown with the U-shaped cradle and wedges in place ready for microcapillaries to be stacked into place.  The U-shaped cradle should be temporarily affixed to the scaffold, with a bit of Kapton tape along the edges, for example.}
\end{center}
\end{figure}

In the subsequent pages, we present the dimensioned drawings for the vacuum chamber segment, the components of the nozzle, and a construction scaffold piece that can aid in construction of the nozzle.  The construction scaffold piece can be used to arrange the U-shaped cradle and wedges as shown in Fig.~\ref{fig:construction_scaffold_assembled}.  The construction scaffolding ensures the ends of the microcapillaries are flush with one face of the U-shaped cradle.  The cover plate can then be attached so that there is no significant gap between the ends of the capillaries and cover plate.  

\clearpage
\pagenumbering{gobble}
\pdfpagewidth=11in \pdfpageheight=17in
\newgeometry{layout=ansibpaper,paperheight=17in,left=0cm,top=0cm,bottom=0cm,right=0cm}
\includegraphics[height=17in]{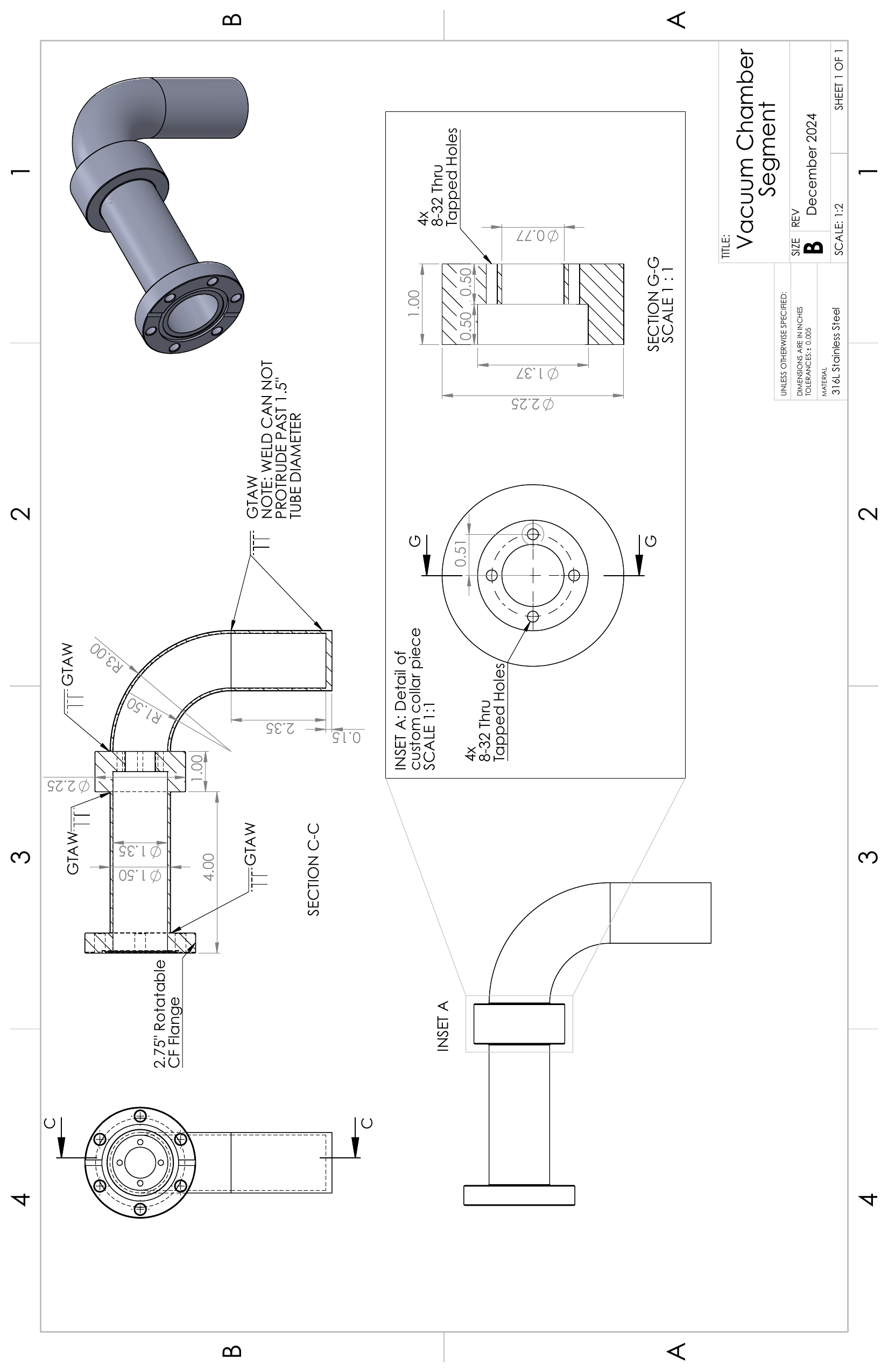}
\includegraphics[height=17in]{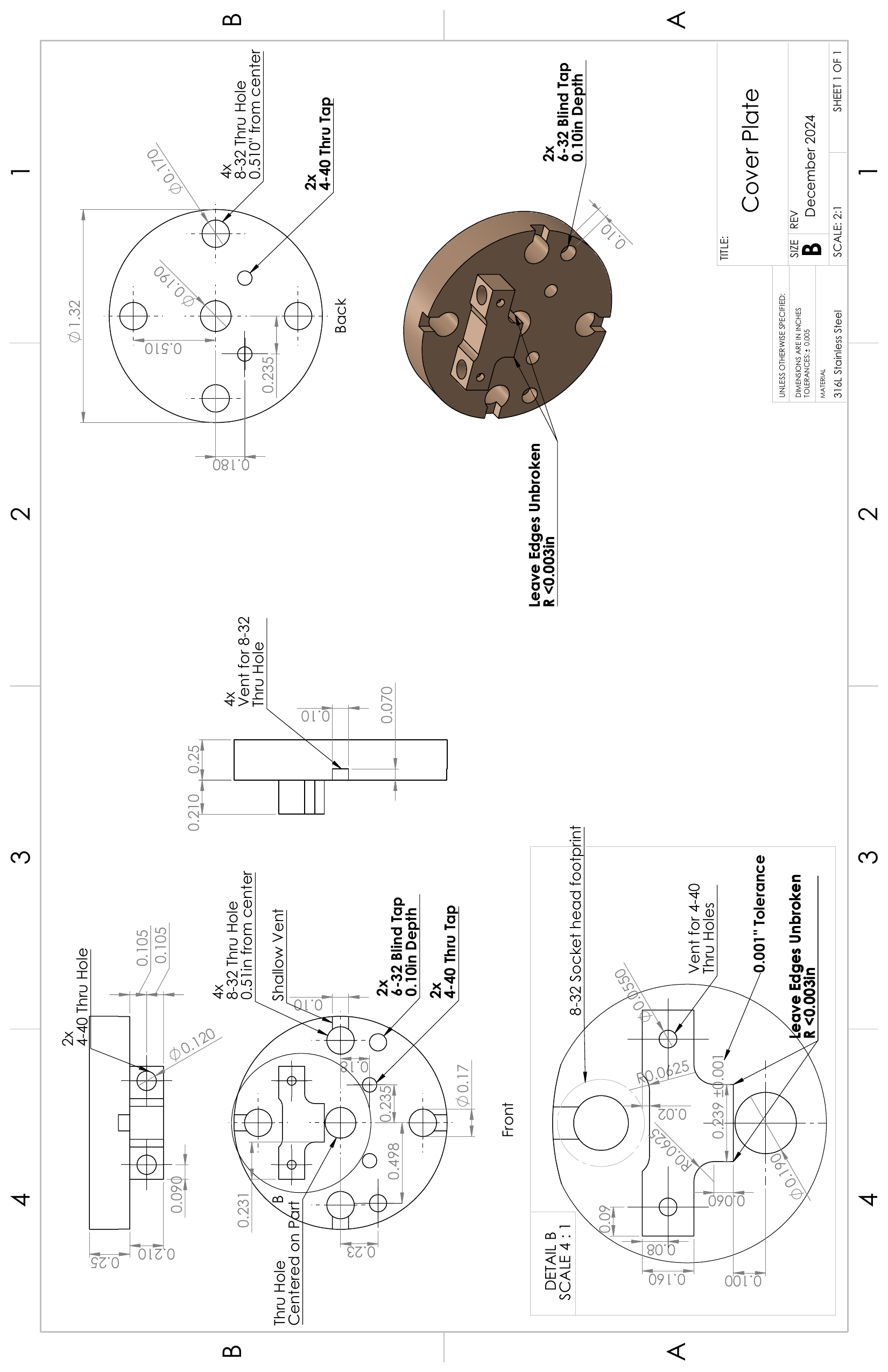}
\includegraphics[height=17in]{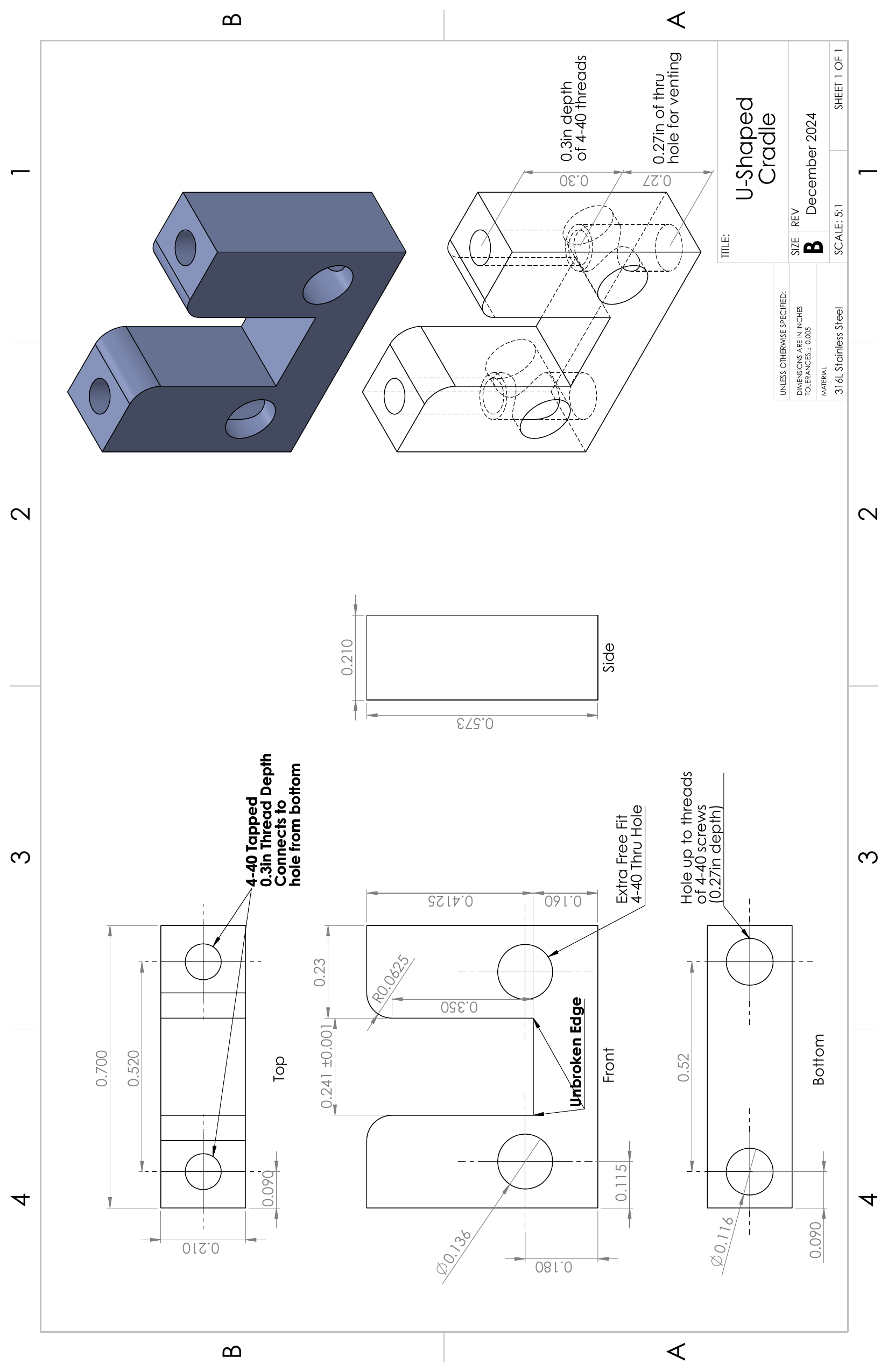}
\includegraphics[height=17in]{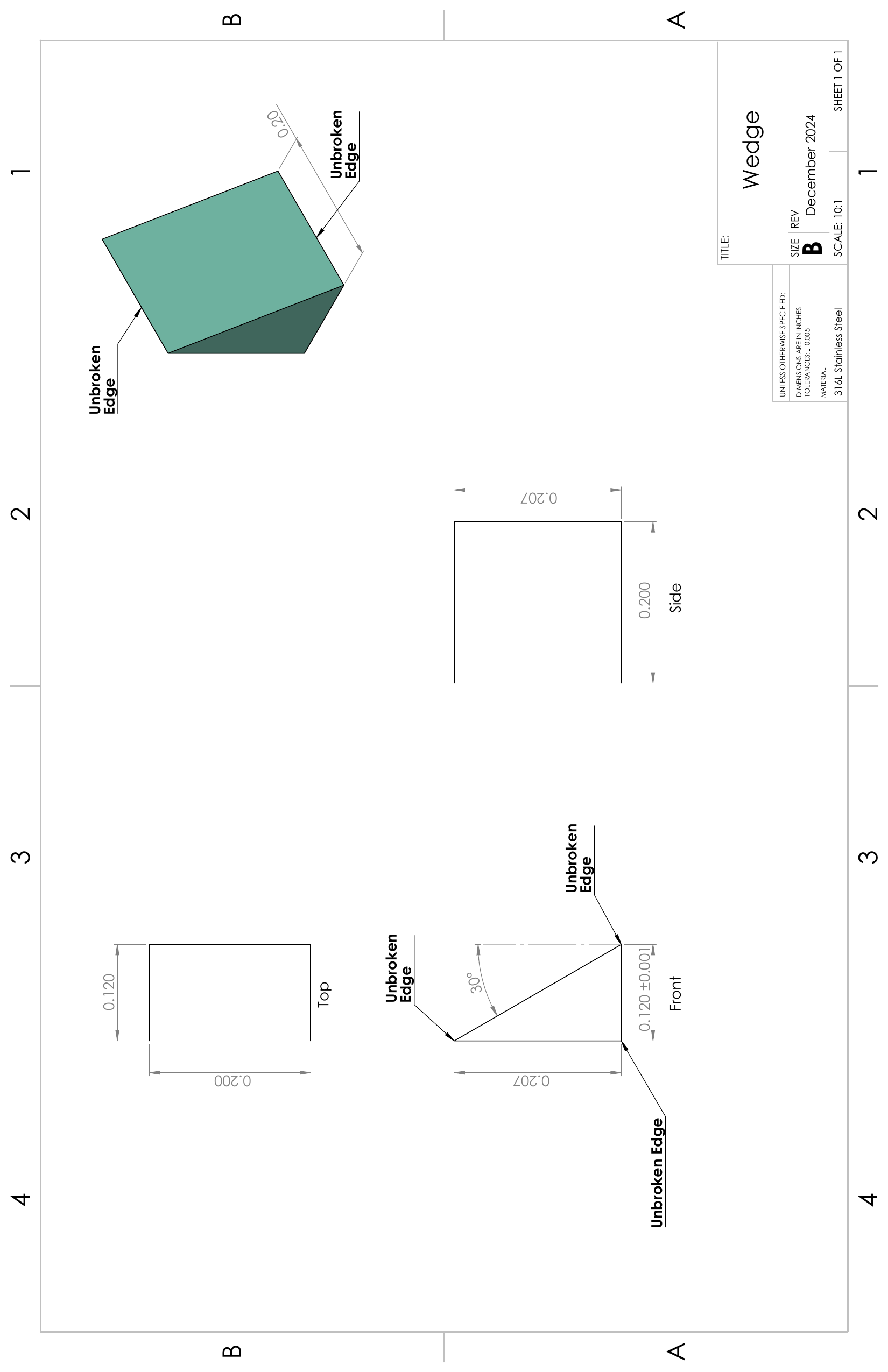}
\includegraphics[height=17in]{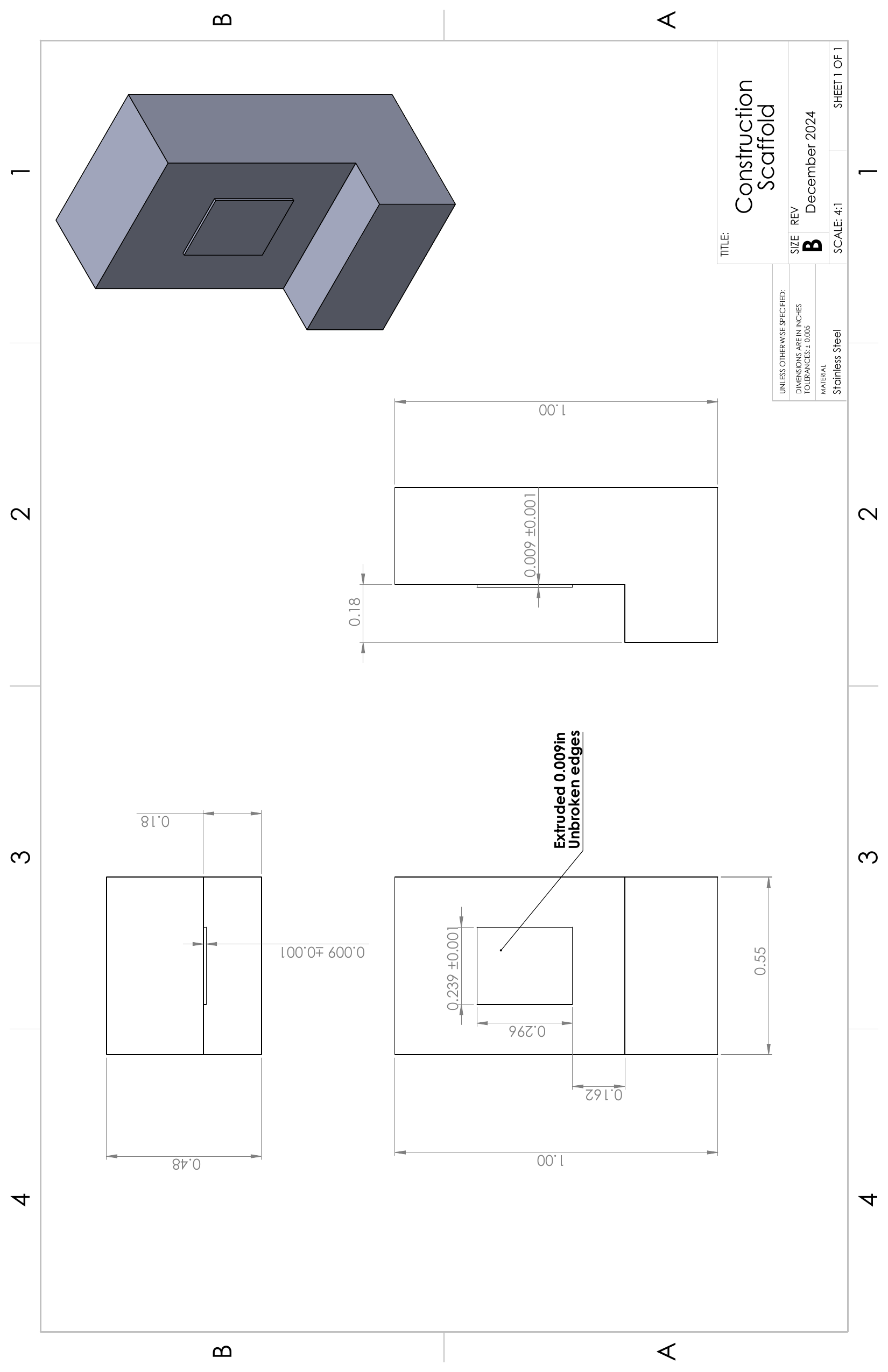}